\documentclass[sigconf,screen,nonacm]{acmart}
\usepackage{tikz}
\usetikzlibrary{shapes,arrows}
\usetikzlibrary{fit}

\AtBeginDocument{%
  \providecommand\BibTeX{{%
    \normalfont B\kern-0.5em{\scshape i\kern-0.25em b}\kern-0.8em\TeX}}}

\begin{document}

\title{Photos Are All You Need for Reciprocal Recommendation in Online Dating}

\author{James Neve}
\affiliation{%
 \institution{Univeristy of Bristol}
 \streetaddress{Tyndall Avenue}
 \city{Bristol}
 \country{United Kingdom}}
\email{james.neve@bristol.ac.uk}

\author{Ryan McConville}
\affiliation{%
 \institution{Univeristy of Bristol}
 \streetaddress{Tyndall Avenue}
 \city{Bristol}
 \country{United Kingdom}}
\email{ryan.mcconville@bristol.ac.uk}


\begin{abstract}
    Recommender Systems are algorithms that predict a user's preference for an item. Reciprocal Recommenders are a subset of recommender systems, where the items in question are people, and the objective is therefore to predict a bidirectional preference relation. They are used in settings such as online dating services and social networks. In particular, images provided by users are a crucial part of user preference, and one that is not exploited much in the literature. We present a novel method of interpreting user image preference history and using this to make recommendations. We train a recurrent neural network to learn a user's preferences and make predictions of reciprocal preference relations that can be used to make recommendations that satisfy both users. We show that our proposed system achieves an F1 score of 0.87 when using only photographs to produce reciprocal recommendations on a large real world online dating dataset. Our system significantly outperforms on the state of the art in both content-based and collaborative filtering systems.
\end{abstract}


\maketitle

\section{Introduction}
\label{sec:introduction}

Recommender Systems (RS) are personalisation tools that are used by online services to recommend items to users \cite{rrs-general-pizzato}. RSs usually make recommendations by computing a preference score between $0$ and $1$ that represents the extent to which a particular user would like a particular item. This is done by using explicit preference information (for instance, a user's profile where they have specified their preference) or implicit preference information, such as a user's purchase history. RSs have become increasingly advanced over the last decade, and most popular online services such as \textit{Amazon} and \textit{Netflix} use them to enhance their users' experience \cite{recsys-general-bell}.

Reciprocal Recommender Systems (RRSs) are a subtype of RSs that recommend people to other people. They are commonly used in online dating and social services. While RSs make recommendations based on a unidirectional preference relation involving an inanimate item, RRSs are inherently more complex in that they must make recommendations based on both sides of a bidirectional preference relation. Applying a conventional recommender system to a reciprocal environment results in recommendations that are only satisfying to one of the two users involved in the interaction. RRS design also involves a number of considerations that are not involved in unidirectional recommendation. For instance, a popular product being repeatedly recommended is not usually a problem, but a popular user appearing in everyone's recommendations often represents a negative experience for that user \cite{rrs-collab-kleinerman}. RRSs are often adapted from RSs, where two unidirectional preferences are computed and then combined into a single preference score that represents the preference of two users for each other.

RSs (and therefore RRSs) are often categorised as content-based or collaborative filtering systems. Content-based systems make recommendations based on a user's preference for specific aspects of an item. These preferences are sometimes explicit, but are more usually inferred implicitly from a user's preference for previous items \cite{rrs-general-aggarwal}. Collaborative filtering algorithms use correlations between multiple users to make recommendations, often by extracting latent factors from a preference matrix of users and items, and inferring preference for those latent factors through historical preferences. Historically, collaborative filtering algorithms have been more effective than content-based algorithms \cite{rrs-general-aggarwal}. However, in content-rich environments, the reverse can be true. Content-based filtering algorithms also tend to be more effective at solving the \textit{Cold-Start Problem} \cite{recsys-cold-lam,recsys-cold-lin}, where it is difficult for the system to make effective recommendations for new users because of the lack of preference history.

Online dating services and social networks often provide a content-rich environment, with users making decisions about whom to express preference to based on a great many factors, including image data, free text profiles and categorical data such as age and job. In particular, image data is extremely important to modern social services, with many such as \textit{Instagram} using images as the basis for interactions. This has been demonstrated by informal research from industry\footnote{https://www.gwern.net/docs/psychology/okcupid/weexperimentonhumanbeings.html}. In this paper, we present a novel recommender system, \textit{Temporal Image-Based Reciprocal Recommender} (TIRR), that uses a Recurrent Neural Network (RNN) to interpret a user's history of preferences for images, and make predictions about their future preferences in order to make recommendations. This is a significant improvement on the only other previous image-based RRS, \textit{ImRec}\cite{rrs-content-neve}, in the sense that it outperforms both ImRec (previously the state of the art in content-based reciprocal recommendation) and also the current state of the art collaborative filtering solutions.

In addition to the advantages in terms of its improvement in the ROC curve on cross-validation, TIRR is also an advance of the field in the sense that it provides a unified system that predicts matches directly, as opposed to two separate predictions of unidirectional preferences followed by an aggregation. There is some doubt as how to combine two unidirectional scores into a single bidirectional score in a way that is fully representative of two users' bidirectional preference for each other; TIRR solves this by predicting the bidirectional relation end to end.

The system was tested using a popular online dating service. We used $200000$ users and approximately $800000$ expressions of preference combined split across train and test sets.

The original contributions of this paper are therefore threefold:

\begin{enumerate}
    \item We present a content-based RRS, TIRR, that makes recommendations based on historical sequences of images utilizing Siamese networks and LSTMs.
    \item Previous RRSs predict two unidirectional preferences and then aggregate them; the end-to-end algorithm detailed in this paper predicts the probability of a match directly.
    \item Based on tests using real-world data, TIRR outperforms not only other content-based RRSs but also the state-of-the-art collaborative filtering RRS.
\end{enumerate}

\section{Related Works}
\label{sec:related-works}

This section contains a review of other academic works that formed the background for this study. This includes works on RRSs, content-based recommendation and on recurrent neural networks for understanding image-based histories.

\subsection{Reciprocal Recommender Systems}

RRSs are recommender systems used for person-to-person matching, in settings such as online dating, social networks \cite{rrs-nodate-he} and job recommendation \cite{rrs-nodate-zheng}. They are complex in the sense that they need to consider the preferences of both sides.

The earliest RRS in the literature is RECON \cite{rrs-content-pizzato}. This is a content-based recommender system designed by Pizzato et al. based on recommendation using categorical data such as age and hobbies. For two users $x$ and $y$, the algorithm calculates the preferences of the two users $P_x$ and $P_y$ as vectors based on their historical preferences for individual attributes. Using these historical preferences, RECON estimates unidirectional preferences $Q_{x, y}$ and $Q_{y, x}$ and combines them using the harmonic mean into a single bidirectional preference relation that represents the projected preference of the two users for each other.

RCF \cite{rrs-collab-xia} extended reciprocal recommendation with an implementation of a collaborative filtering system. RCF uses nearest-neighbour-based recommendation, where for candidate users $x$ and $y$ it calculates the similarity between $x$ and the other users that have liked $y$ and vice-versa. RCF demonstrated improvements in a number of areas over RECON, and was at the time considered the best in class reciprocal recommender system.

Subsequently, a number of systems have made improvements to both collaborative and content-based systems, in addition to designing hybrid systems that exploit the best of both subtypes. For example, Kleinermann et al. improved on RCF by reducing the bias of user popularity on recommendations \cite{rrs-collab-kleinerman}. ImRec demonstrated that image-based recommendation was more effective than recommendation based on categorical profile data \cite{rrs-content-neve}.

\subsection{Content-Based Recommender Systems}

Content-based recommender systems make recommendations based on users' preferences for specific content on a service. This might be structured content such as the category of an item, or it might be unstructured content such as images and free text description.

Recommendations based on unstructured data appear most often in news recommendation \cite{recsys-content-lang, recsys-content-meteren}. This is a rich area for research because news articles are often written with set structures that make them easier to process, and also because of concerns about serendipity and recommendations reinforcing existing echo chambers. Papers such as \cite{recsys-content-bansal} demonstrate the capacity for deep learning systems based on freetext information to make effective recommendations.

Examples of content-based recommender systems basing their results on images is much less common. Lei et al. used user preferences to train a model based on ImageNet \cite{ml-images-deng} that predicted user preference for one of two images \cite{recsys-content-lei}. This trains a network to map both users and images into the same space by generating embeddings for both, with images that the user preferred being close to the user in the space, and images the user did not like being further away. User preference for subsequent images can then be predicted by relative distance from the user.

Another example is \textit{DeepStyle} \cite{recsys-content-liu-2}, which uses a Siamese Network to predict user preference for clothes based on images. DeepStyle uses pairs of positive and negative samples with user preference as the output to differentiate between the two images. This can then be used to make predictions about whether a user might like a new image by comparing it to an existing liked image.

\subsection{Recurrent Neural Networks}

This paper uses Recurrent Neural Networks (RNNs) to interpret time series data for recommendation. RNNs contain loops, which feed the output of a network back into current neurons. This means that they implement a concept of \textit{memory}: they store computed results, and these results have an impact on subsequent predictions. Each step therefore incorporates information from the previous steps into the prediction.

Standard RNNs are particularly good at processing short sequences, but their memory is \textit{short-term memory}: when training them using longer sequences, the early items in the sequence have very little impact on the final prediction. This is known as the \textit{vanishing gradient problem} \cite{ml-general-hochreiter}.This also exists in deep neural networks, where early layers learn very slowly when trained with backpropagation.

Various architectures have been proposed to overcome this limitation. One that has been particularly successful in allowing RNNs to hold and use information for longer is the \textit{Long Short-Term Memory Network} (LSTM) \cite{ml-general-hochreiter-2}. A LSTM uses a \textit{forget gate} comprised of a Sigmoid function that determines whether information is kept or not: a value close to $0$ results in the information being forgotten by the network, whereas closer to $1$ results in the information being stored. This allows for much longer sequences to be processed, which is particularly useful in the field of recommendation, where long sequences of user behaviour are common. RNNs have been used successfully in recommender systems to incorporate time series data into recommendations \cite{recsys-general-twardowski, recsys-general-wu}, but not in reciprocal recommender systems. 

\section{Methodology}

In this section, we describe a model that produces predictions about user preferences based on the RNN interpretation of user history. The RNN-based model uses a pre-trained siamese network at its core, so we describe that independently first, and then its use in the context of the RNN.

\subsection{Problem Formulation}
\label{sec:probfor}

The online dating service we used currently only supports heterosexual relationships. We can therefore assume that for a set of users of one gender $X = \{x_1, x_2, \ldots, x_{|X|}\}$ there is a set of candidate users for recommendation $Y = \{y_1, y_2, \ldots, y_{|Y|}\}$. A user may have an ordered history of preference expressions for users of length $n$, for example, $Sx = \{Sx^{y_i}_{t_0}, Sx^{y_j}_{t_1}, ^{y_k}_{t_m}, \ldots, Sx^{y_l}_{t_n}\}$ where $Sx^{y_i}_{t_m} \in Y$ represents the expression of positive or negative preference of user $x$ for the user $y_i$ at time $t_m$.

In our reciprocal system, our objective is to estimate $R^{x,y}$, the reciprocal  preference score that represents the projected degree of preference of two users for each other. We consider that $R^{x, y}$ is a function of the historical preferences of $x$ and $y$ as well as the two users themselves, and train a model to predict it using all of this information:

\begin{equation}
    R^{x, y} = f(S^x, S^y, x, y; \theta)
\end{equation}

Where $\theta$ represents the parameters of the model. Note that contrary to most previous approaches to RRSs, our approach trains a single model to predict reciprocal preference using all of the information, as opposed to combining the results of two models predicting unidirectional preference. Also note that the reciprocal preference is symmetrical i.e. $R^{x, y} = R^{y, x}$.

\subsection{Service \& Data}
\label{sec:serdat}

The data for our model was provided by a popular online dating service with several million registered users. On this service, the user experience is streamlined so that everyone goes through the same process of interaction.

A user $x$ finds other users by searching, or by viewing recommendations on a list page. From the list page, they can view profiles with images, text and categorical data. If $x$ finds a user $y$ that they want to interact with, they can send a $Like$. In our algorithm, this $Like$ is used as a unidirectional indicator of preference.

User $y$ can choose whether or not to reciprocate this $Like$. If they do reciprocate, this is considered a $Match$; if not it is considered a $Dislike$. These represent bidirectional indicators of preference or negative preference respectively. Users who have $Matched$ can subsequently message each other, and potentially agree to meet in person.

As we wanted to focus on an algorithm that measured personal attractiveness of users for each other, we made the decision to exclude images from the dataset that did not include user faces using automatic face detection. It is common for deep learning based on faces to also include cropping an affine transformation of features, but preliminary experiments showed that this did not improve our results.

We also made a number of exclusions in order to increase the reliability of the dataset. We excluded users who had been removed from the service for any reason (often these users are not using the service correctly, which implies that they are not expressing preferences based on their own intuition). Although for privacy reasons we are unable to release the dataset used in our experiments, we do hope that the algorithm will be reproduced on other services.

\subsection{Siamese Network Unidirectional User Preference Learning}

\begin{figure}[!htb]
    \center{\includegraphics[width=0.45\textwidth]
    {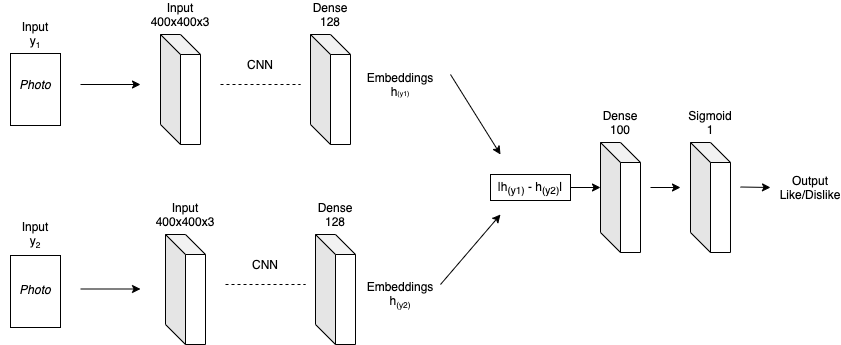}}
    \Description[Siamese Network Architecture]{A representation of the Siamese Network.}
    \caption{\label{fig:siamese-network} The architecture of the Siamese network used to learn unidirectional user preferences, which forms a component of TIRR.}
\end{figure}

In this section, we briefly describe the Siamese network \cite{ml-images-koch} we used to learn unidirectional user preferences, a building block of our proposed method, originally included as a component of \textit{ImRec} \cite{rrs-content-neve}. We will utilize this in a novel way to \textit{ImRec} to demonstrate superior performance.

\subsubsection{Siamese Network Concept}

Siamese networks are commonly used in object recognition \cite{ml-images-vinyals} and tracking \cite{ml-images-bertinetto}, where they have excelled for face verification in scenarios with relatively little training data, known as \textit{one-shot} or \textit{few-shot learning}. As shown in Figure \ref{fig:siamese-network}, a Siamese network consists of two symmetrical CNNs with shared weights, and a loss function based on the outputs of these two networks and a ground truth label.

The network is trained from tuples of the form $(y_a, y_n, y_p)$ from $Y$ where $y_a$ and $y_p$ are photos of $y$ that have been \textit{Liked} by $x$ and $y_n$ is a photo of $y$ that has been \textit{Disliked} by $x$. Using $y_a$ as an anchor, two pairs are made from the tuple; $(y_a, y_p)$ is a positive pair where the expected output is $1.0$ and $(y_a, y_n)$ as the negative pair where the expected output is $0.0$. From these pairs, the network is trained to differentiate between a Liked and a Disliked image, given another Liked image.

The key to this is that the network uses shared weight parameters $W$ for the training and inference process. We map $y_1$ and $y_2$ to $h_{y1}$ and $h_{y2}$ using $W$, which are two points in a 128 dimensional space. We calculate the difference between the two points as follows:

\begin{equation}
    D_W(y_1, y_2) = |h_{y1} - h_{y2}|
\end{equation}

Relating the Siamese network back to our original problem formulation in Section \ref{sec:probfor}, this gives us a basis for estimating a unidirectional preference relation $P^{x,y}$ based on two images, one image from $x$'s preference history $S^x_k$, and the current user $y$, solving the problem:

\begin{equation}
    P^{x,y} = g(S^x_k, y; \theta)
\end{equation}

\subsubsection{Network Layers}

\begin{table}[h]
\begin{tabular}{|l|l|l|l|l|}
\hline
\textbf{Layer}  & \textbf{Size-in} & \textbf{Size-out} & \textbf{Kernel} & \textbf{Param} \\ \hline
input &                    & 100x100x3                 &  & 0                  \\  \hline
conv1 & 100x100x3         & 100x100x3                & 7x7x3 & 444            \\  \hline
maxpooling1 & 100x100x3         & 34x34x3                & 3x3 &             \\  \hline
normalization1 & 34x34x3         & 34x34x3                &  & 12            \\  \hline
conv2 & 34x34x3         & 34x34x64                & 3x3x64  & 1792            \\  \hline
maxpooling2 & 12x12x64         & 12x12x64                & 3x3 &             \\  \hline
normalization2 & 12x12x64         & 12x12x64                &  & 256            \\  \hline
conv3 & 34x34x3         & 12x12x192                & 2x2x192  & 49344            \\  \hline
maxpooling3 & 12x12x64         & 4x4x192                & 3x3 &             \\  \hline
conv4 & 4x4x192         & 4x4x384                & 2x2x384  & 295296            \\  \hline
maxpooling4 & 4x4x384         & 2x2x384                & 3x3 &             \\  \hline
conv5 & 2x2x384         & 2x2x256                & 1x1x256  & 98560            \\  \hline
conv6 & 2x2x256         & 2x2x256                & 3x3x256  & 590080            \\  \hline
maxpooling5 & 2x2x256         & 1x1x256                & 3x3 &             \\  \hline
flatten & 1x1x256         & 256                &  &             \\  \hline
dense1 & 256         & 256                &  &  65792           \\  \hline
dense2 & 256         & 128                &  &  32896           \\  \hline
\end{tabular}
\caption{The layers of the CNN used as the symmetrical part of the Siamese Network.}
\label{tbl:cnn-table}
\vspace{-4mm}
\end{table}

Table \ref{tbl:cnn-table} shows the architecture of the CNN that makes up the two symmetrical branches of the Siamese network. The small convolution kernels used have been shown to effectively identify facial features in deep convolutional networks \cite{ml-images-parkhi}. The network was trained using an Adam optimiser, with a learning rate of $0.0001$.

The output of the network is a value between $0$ and $1$ expressed by a Sigmoid function, representing whether $x$ is more likely to \textit{Like} or \textit{Dislike} $y$ based on the two images. In the next section, we extend this to use the whole of $x$'s preference history $S_x$ and show how this becomes an even more effective predictor of preference.

\subsubsection{Loss Function}

The Siamese network was trained using \textit{binary crossentropy}. This is a standard loss function used in training neural networks, the formula for which is given below. In the following equation, $Y$ is the binary variable representing Like and Dislike, $D_W(y_1, y_2)$ is the embedded distance between two images, $g$ is a neural network and $g(D_W(y_1, y_2))$ is the predicted probability of $D_W(y_1, y_2)$ resulting in a Like.

\begin{equation}
   L(y_1, y_2) = -{(Y\log(g(D_W(y_1, y_2))) + 
   (1 - Y)\log(1 - g(D_W(y_1, y_2))))}
    \label{eq:bce}
\end{equation}

Binary crossentropy was shown experimentally to result in the highest effectiveness metrics for the network.

We note that it is also common to train Siamese networks using a \textit{contrastive loss} function, which uses a \textit{margin} $m$ to increase the network's error when it misclassifies two very similar images. In most situations where a Siamese network is applied, such as face detection, misclassifying two very similar images is as incorrect as misclassifying two very different images. However, this is not true in our application, where user preferences are not necessarily categorical, and similar images are more likely to be liked by a user than very different images. This explains why binary crossentropy may have been more effective in our tests.

The contrastive loss function is defined below, where the terminology used is the same as in equation \ref{eq:bce} and $m$ is the margin:

\begin{equation}
     L(y_1, y_2) = (1 - Y)\frac{1}{2}(D_W(y_1, y_2))^2 + Y\frac{1}{2}(max(0, m - D_W(y_1, y_2))^2
\end{equation}

\subsection{Incorporating RNNs for Learning User Preference History}

\begin{figure*}[!htb]
    \center{\includegraphics[width=0.9\textwidth]
    {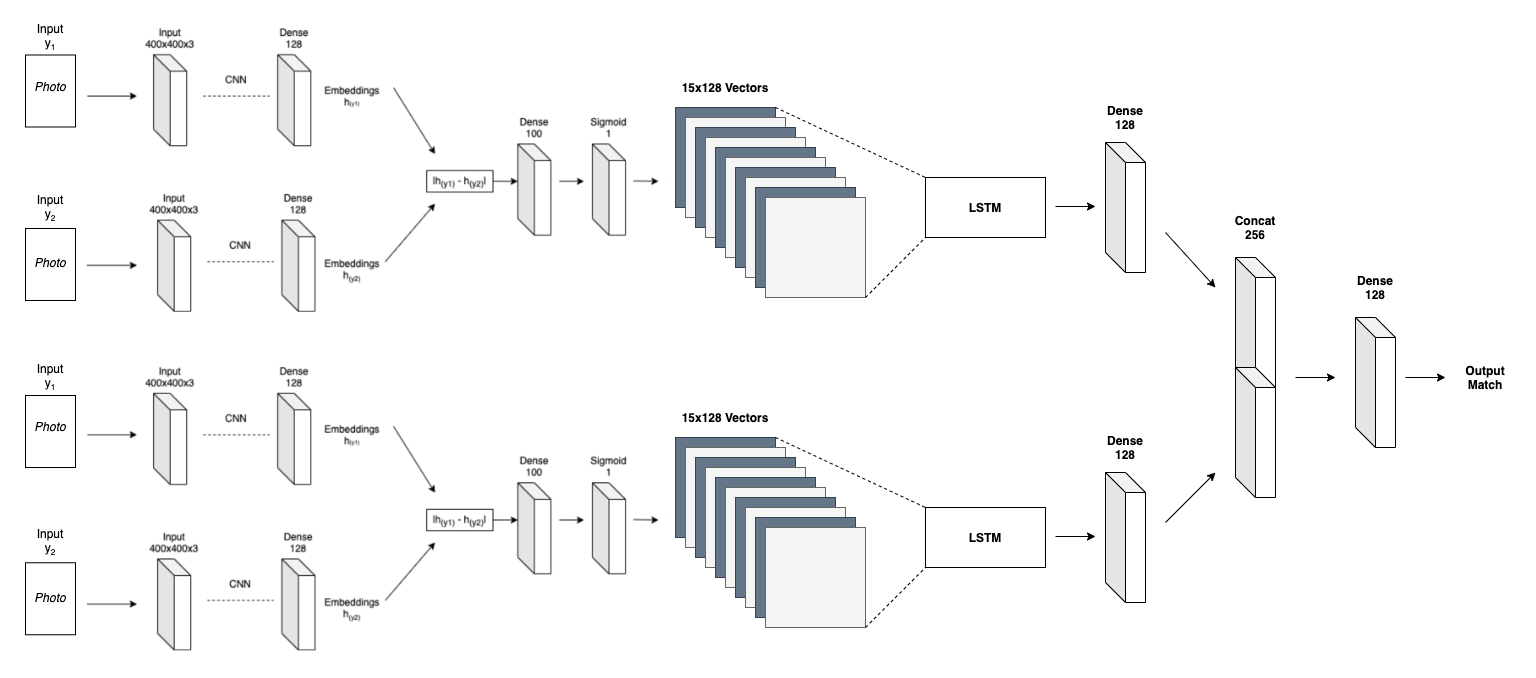}}
    \Description[LSTM Network Architecture]{A representation of the LSTM trained with outputs from pre-trained Siamese Networks.}
    \caption{\label{fig:lstm-siamese} TIRR: the architecture to predict matches using an LSTM to interpret historical preference data on user photos.}
\end{figure*}

The Siamese network described above, when trained on unidirectional preference, is an effective if elementary model. In this section, we describe the RNN we use to interpret the user history based on the results of the Siamese network.

The output of the Siamese network is a point in 128-dimensional space that represents the preference of a user $x$ for an image $y_k$ based on comparison with the anchor image $y_a$. Based on initial experimental work, we chose an LSTM-based RNN architecture to interpret the time series of images. The \textit{forget gate} of the LSTM is particularly intuitive in this case. For a state $s_t$ at time $t$, a forget gate described by $f_t$, a write gate $i_t$ and a candidate write $\tilde{s}_t$ derived from the input and the previous state, the next state is described by the equation:

\begin{equation}
    s_t = f_t \odot s_{t-1} + i_t \odot \tilde{s}_t
\end{equation}

We might intuitively expect that preferences expressed by users would change over time, and the forget behaviour of the LSTM allows us to model this, with the input for the state $s_t$ of the LSTM modelling the preferences of user $x$ being the user $S^x_t$, and the final input at $s_{|S^x| + 1}$ being the user $y$ whom we wish to estimate $x$'s preference for.

The LSTM is visualised in Figure \ref{fig:lstm-siamese}. Because users have variable length preference histories, we fill the histories of users with shorter histories with dummy images and use a masking layer to filter them. The LSTM and subsequent dense neural network form a representation in 256-dimensional space of the user's preference as a time series.

\begin{table}[h]
\begin{tabular}{|l|l|l|l|l|}
\hline
\textbf{Layer}  & \textbf{Size-in} & \textbf{Size-out} & \textbf{Kernel} & \textbf{Param} \\ \hline
input &                    & 128x15                 &  & 0                  \\  \hline
LSTM &                    & 128x15                 & 128 & 128                  \\  \hline
dense1 &                    & 128                 & 128 &                   \\  \hline
concat &                    & 128x2                 & 256 &                   \\ \hline
dense2 &                    & 256                 & 128 & 128                  \\  \hline
output &                    & 128                 & 1 &                   \\  \hline
\end{tabular}
\caption{The layers of TIRR following the mapping of images into 128-dimensional space by the pre-trained Siamese network}
\label{tbl:lstm-table}
\vspace{-4mm}
\end{table}

Specifically, the network consists of an input layer, which accepts a maximum of $15$ outputs from Siamese networks in 256-dimensional space concatenated together. Experiments determined that more than this did not significantly alter the performance of the network. The layers are described in Table \ref{tbl:lstm-table}. If a user has fewer preferences expressed than this, the earlier images are filled with zeroes, and the network learns to interpret this as dummy data. Following the LSTM, the network consists of a single dense layer of 128 neurons, and then a dropout layer with a dropout rate of $0.4$. The network was trained with an Adam optimiser with a learning rate of $0.0001$. 

\subsection{Training and Match Prediction}

\tikzstyle{decision} = [diamond, draw, fill=blue!20, 
    text width=4.5em, text badly centered, node distance=3cm, inner sep=0pt]
\tikzstyle{block} = [rectangle, draw, fill=blue!20, 
    text width=5em, text centered, rounded corners, minimum height=4em]
\tikzstyle{line} = [draw, -latex']
\tikzstyle{cloud} = [draw, ellipse,fill=red!20, node distance=3cm,
    minimum height=2em]
    
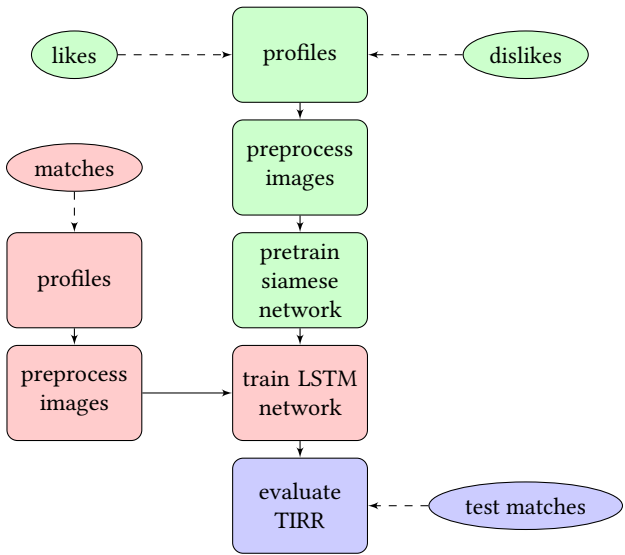
\begin{figure}
    \begin{tikzpicture}[node distance = 1.5cm, auto]
        \node [block, fill=green!20] (profiles) {profiles};
        \node [cloud, left of=profiles, fill=green!20] (likes) {likes};
        \node [cloud, right of=profiles, fill=green!20] (dislikes) {dislikes};
        \node [block, below of=profiles, fill=green!20] (preprocess) {preprocess images};
        \node [block, below of=preprocess, fill=green!20] (siamese) {pretrain siamese network};
        \node [cloud, left of=preprocess, fill=red!20] (matches) {matches};
        \node [block, below of=matches, fill=red!20] (profilesmatch) {profiles};
        \node [block, below of=profilesmatch, fill=red!20] (preprocessmatch) {preprocess images};
        \node [block, below of=siamese, fill=red!20] (lstm) {train LSTM network};
        \node [block, below of=lstm, fill=blue!20] (validate) {evaluate TIRR};
        \node [cloud, right of=validate, fill=blue!20] (matchestest) {test matches};

        \path [line] (profiles) -- (preprocess);
        \path [line] (preprocess) -- (siamese);
        \path [line] (profilesmatch) -- (preprocessmatch);
        \path [line] (siamese) -- (lstm);
        \path [line] (preprocessmatch) -- (lstm);
        \path [line] (lstm) -- (validate);
        \path [line,dashed] (likes) -- (profiles);
        \path [line,dashed] (dislikes) -- (profiles);
        \path [line,dashed] (matches) -- (profilesmatch);
        \path [line,dashed] (matchestest) -- (validate);
        
    \end{tikzpicture}
    \Description[Training Flowchart]{A flowchart describing how the Siamese Network and subsequent LSTM network are trained and validated}
    \caption{\label{fig:training-flowchart} The process by which TIRR is trained. Three independent datasets used represented by different colours.}
\end{figure}

This section describes training the network to predict matches between two users. As described in Section \ref{sec:serdat}, our objective is to differentiate between interactions consisting of bidirectional expressions of preference, \textit{Matches}, and unidirectional expressions of negative preference, \textit{Dislikes}.

The full training process is visualised in Figure \ref{fig:training-flowchart}. Our experiments determined that the network trained extremely slowly when trained in its full form from an initial randomised state, and we therefore pre-trained the Siamese network segment of the network using one dataset, shown in green. The subsequent training of the full system on matches was done using a separate dataset, shown in red. The final evaluation was done using a third dataset, shown in blue. In addition, Neve et al. demonstrated that the Siamese Network training was more effective when two networks were trained separately on male and female data \cite{rrs-content-neve}. As the service providing our data currently only supports heterosexual dating, this split does not decrease the usefulness of the application in this case.

Training for the Siamese networks were based on $500000$ triplets $(y_a, y_p, y_n)$ sampled from $250000$ users split evenly over male and female images. Images were cropped and centered on the faces of users before training. Other methods of preprocessing such as affine transformations, which have been shown to improve the predictive power of other networks \cite{ml-images-lewenberg} did not have any impact on performance. The Siamese networks were trained to predict unidirectional preferences i.e. $y_p$ was an image $x$ had \textit{Liked} (but not necessarily with reciprocity) and $y_n$ was an image  $x$ had \textit{Disliked}.

Following convergence of the Siamese network, the LSTM network was trained based on the preference histories of $100000$ users to predict \textit{Matches} and \textit{Like-Dislike Tuples}. This dataset was separate from the dataset used to train the Siamese network. Histories were capped at one year, because of concerns that changes to the service's design and search algorithm over time might have an effect on user preferences. They were also capped to a maximum of $15$ preferences, because initial experiments showed that longer sequences did not improve accuracy, and because some outlier users express thousands of preferences, which results in an unreasonable increase in training and prediction times.

Finally, the LSTM was validated on a separate dataset of $20000$ \textit{Matches} and \textit{Like-Dislike Tuples}. There was no overlap in preference expression between the three datasets. There was overlap between the users contained in these datasets, but as in a real-world situation the system would be trained based on users on the service and subsequently used to make predictions for those users in addition to new users, testing in this way is valid and representative.

\section{Results \& Evaluation}

In this section, we present the results for TIRR compared to the current state of the art in both content-based and collaborative filtering.

\subsection{Evaluation Metrics for Reciprocal Environments}

RRSs generally use similar metrics for success as standard machine learning models: evaluation via the \textit{ROC Curve} and the related metrics \textit{Precision}, \textit{Recall} and their combined metric \textit{F1 Score}. However, because of the requirement for reciprocal success, their definitions are a little different in this scenario, so we present them here as defined by Pizzato et al. in \cite{rrs-general-pizzato}. In the following equations, $R$ is the set of recommended users, $RL$ is the set of recommended users who matched with each other, and $RN$ is the set of recommended users where at least one expressed negative preference for the other.

\begin{equation}
Precision = \frac{|RL|}{|RL| + |RN|}
\end{equation}

\begin{equation}
Recall = \frac{|RL|}{|R|}
\end{equation}

\begin{equation}
F1 = \frac{2 * Precision * Recall}{Precision + Recall}
\end{equation}

As the models predict a value between $0.0$ and $1.0$ that represents the strength of the mutual preference relation, the ROC curves in this section are drawn by moving a threshold between these two values and plotting the true and false positive rates.

\subsection{Siamese Network Results}

We first describe the results generated by the pretrained Siamese network. This network provides a basis for the main user preference prediction model, as the output embeddings from this network provide an input for the RNN.

\begin{figure}[!htb]
    \center{\includegraphics[width=0.45\textwidth]
    {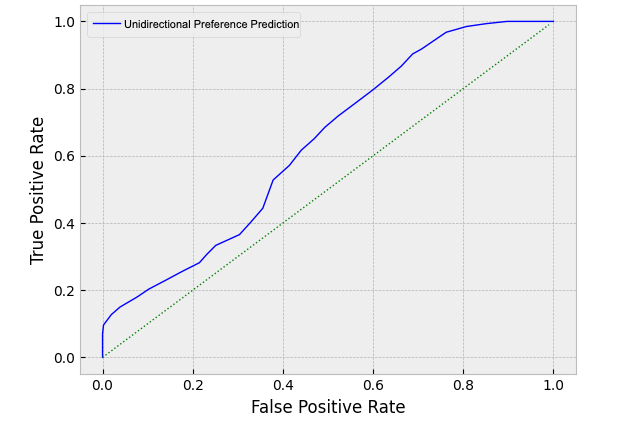}}
    \Description[Siamese Network ROC Curve]{The ROC curve for the Siamese network pretrained to differentiate between two Liked images and a Liked and a Disliked image.}
    \caption{\label{fig:roc-siamese} Pretrained Siamese Network ROC curve. This forms a building block of both ImRec and our proposed TIRR model. }
\end{figure}

The ROC curve for the Siamese network is displayed in Figure \ref{fig:roc-siamese} as the blue line (the green dotted line is the $1$-$0$ reference line). This curve was drawn based on a test set of $20000$ samples not in the original training dataset. In general the network is capable of differentiating between a single \textit{Liked} image and a single \textit{Disliked} image based on an anchor image. The curve itself is slightly erratic, but this is not entirely unexpected: a single anchor image is unlikely to enough information to differentiate between positive and negative preference.

As the model by itself is not directly the source of the recommendations, it would not be appropriate to compare it to other recommender systems. For this reason, we present this model without a point of comparison. However, in Section \ref{sec:rnn_comp} we will compare two approaches that use this model as a building block for reciprocal recommendation. 

\begin{figure}[!htb]
    \center{\includegraphics[width=0.45\textwidth]
    {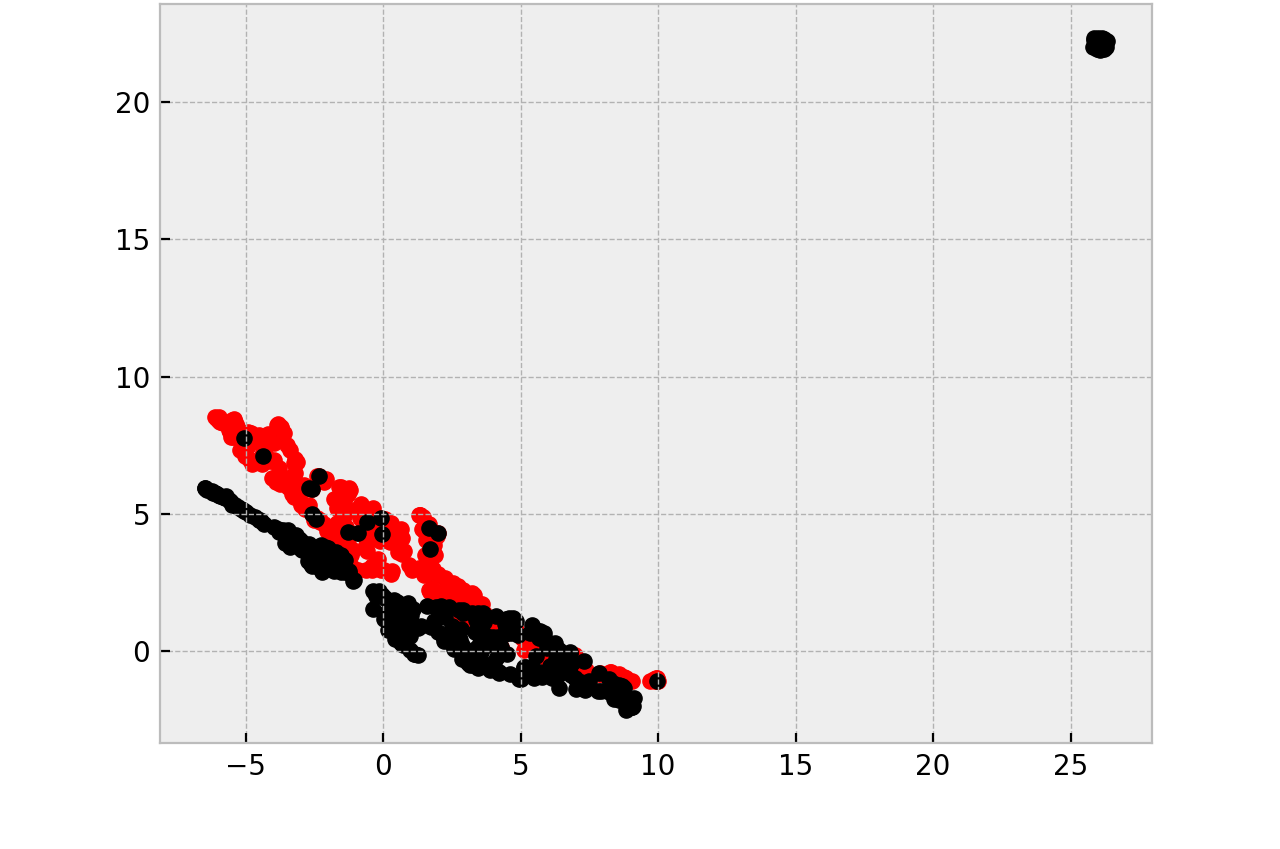}}
    \Description[Siamese Network Embeddings]{Embeddings for the pretrained Siamese Network, for 500 random positive and negative samples.}
    \caption{\label{fig:embeddings-siamese} UMAP embeddings of the pretrained Siamese network forming part of TIRR. The red points represent \textit{Liked} images while black points represent \textit{Disliked} images.}
\end{figure}

As the embeddings in 128-dimensional space from the output of the Siamese network form the input of the RNN. It is therefore useful to visualise these embeddings. In order to do this, we use \textit{Uniform Manifold Approximation and Projection for Dimensionality Reduction} (UMAP) to reduce the 128-dimensional vectors to two-dimensional vectors for visualisation. This visualisation is displayed in Figure \ref{fig:embeddings-siamese}.

In this visualisation, the black datapoints represent \textit{Disliked} images and the red datapoints represent \textit{Liked} images. It is clear from the visualisation that the embeddings are separable to some extent, even in two dimensions. The anomalous black cluster in the top right of the image represents heavily distorted or very poor quality images, or images misclassified by the face detection algorithm (i.e. images that do not contain a face). These tend to be almost universally \textit{Disliked}.

\subsection{TIRR vs Content-Based Algorithms}
\label{sec:rnn_comp}
As described in Section \ref{sec:introduction}, recommender systems are divided into content-based algorithms and collaborative filtering algorithms.

\begin{figure}[!htb]
    \center{\includegraphics[width=0.45\textwidth]
    {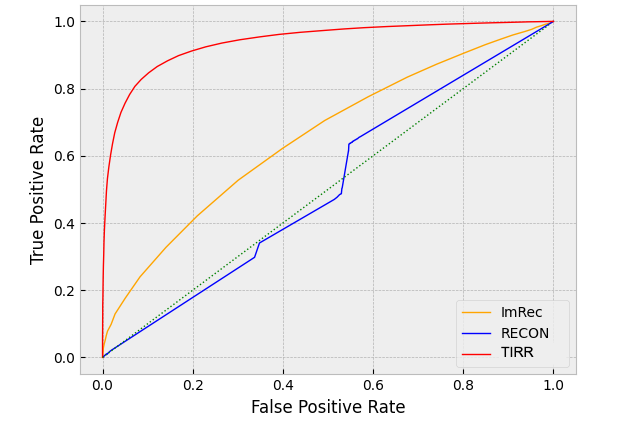}}
    \Description[Content-Based Algorithm ROC Curves]{ROC Curves for the content-based RRS Algorithms RECON, ImRec and TIRR.}
    \caption{\label{fig:content-roc-curves} Content Based Algorithm ROC Curves demonstrating the significant improvement in AUC with TIRR.}
\end{figure}

Figure \ref{fig:content-roc-curves} displays a comparison of TIRR with other content-based algorithms. As described in Section \ref{sec:related-works}, \textit{RECON} \cite{rrs-content-pizzato} is a an algorithm that identifies a user's implicit preferences for categorical data, and \textit{ImRec} \cite{rrs-content-neve} is an algorithm that uses images to make predictions without the RNN-based component of TIRR, instead using a Random Forest and aggregation function.

\textit{RECON} struggled to generate effective recommendations on our dataset. As RECON was also evaluated on a private dataset, without comparing the datasets directly, it is difficult to establish why this is, but one possibility is that modern dating services place a higher emphasis on visual content than services did ten years ago, at the time \textit{RECON} was developed. \textit{ImRec} performs better than \textit{RECON}, but performs significantly worse than our proposed method \textit{TIRR}. The key difference between \textit{TIRR} and \textit{ImRec} is the RNN-based process that allows \textit{TIRR} to interpret historical and time-series data in order to make predictions, whereas \textit{ImRec} treats user preferences in a global way, with no ability to capture individual users preferences.

\begin{table}[h]
\begin{tabular}{|l|l|l|l|l|}
\hline
\textbf{Algorithm} & \textbf{F1 Score} & \textbf{Precision} & \textbf{Recall} & \textbf{AUC} \\ \hline
\textit{RECON} & 0.61  & 0.56 & 0.68 & 0.51 \\ \hline
\textit{ImRec} & 0.71  & 0.60 & 0.88 & 0.65 \\ \hline
\textit{TIRR} & 0.87 & 0.86 & 0.88 & 0.91 \\ \hline
\end{tabular}
\caption{Results based on best F1 score for content-based algorithms. Here we can see that our proposed method TIRR significantly outperforms the other approaches.}
\label{tbl:content-f1-score}
\vspace{-4mm}
\end{table}

The AUC and maximum F1 score for the three algorithms is described in Table \ref{tbl:content-f1-score}. The scores are based on the threshold that gave the best F1 score in the training set, used in the test set. We consider that this significant improvement of our proposed method \textit{TIRR} derives from the ability of our algorithm to interpret a user's history of preferences for images over time, and take account of a user's potentially shifting preferences, whereas \text{Imrec} provides a global model across all users without distinguishing more than one preference per user at a time, and \textit{RECON} doesn't make use of images at all.

The table also lists the precision and recall at the points where the best F1 score was recorded. While F1 is an excellent measure of overall performance of an algorithm, the individual precision and recall numbers and their balance are particularly important in RS research because precision tends to influence the trust users have in the RS, which in turn affects their use of it \cite{recsys-eval-herlocker}. It is noteworthy that while \textit{ImRec} was relatively successful at predicting which image a user would like, its precision was relatively low in comparison with other algorithms, whereas \textit{TIRR} has very high precision, and is therefore more likely to be trusted and used.

\subsection{TIRR vs Collaborative Filtering}

In addition to comparing \textit{TIRR} to other content-based RRSs, we also ran tests comparing it to the current best-in-class collaborative filtering algorithms, \textit{RCF} and \textit{LFRR}.

\begin{figure}[!htb]
    \center{\includegraphics[width=0.45\textwidth]
    {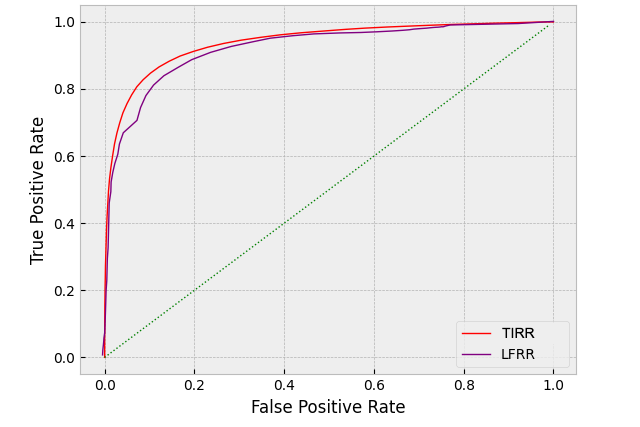}}
    \Description[Collaborative Filtering Algorithm ROC Curves]{ROC Curves for the collaborative filtering RRS Algorithms LFRR, RCF and TIRR.}
    \caption{\label{fig:collaborative-roc-curves} ROC Curves showing the performance of the content-based TIRR against the current state of the art collaborative filtering algorithm LFRR.}
\end{figure}

\textit{LFRR} is a collaborative filtering algorithm based on latent factor models trained by stochastic gradient descent, and \textit{RCF} is a neighbourhood-based collaborative filtering algorithm. \textit{TIRR} outperformed both of these algorithms on our test dataset, although by a slimmer margin than its lead on current content-based filtering algorithms. Nonetheless, this represents a significant advancement in the field of reciprocal recommendation, as in services where images prominently used, our algorithm is likely to be more effective than current collaborative filtering methods.

\begin{table}[h]
\begin{tabular}{|l|l|l|l|l|}
\hline
\textbf{Algorithm} & \textbf{F1 Score} & \textbf{Precision} & \textbf{Recall} & \textbf{AUC} \\ \hline
\textit{LFRR} & 0.86  & 0.86 & 0.85 & 0.90 \\ \hline
\textit{TIRR} & 0.87 & 0.86 & 0.88 & 0.91 \\ \hline
\end{tabular}
\caption{Results based on best F1 score for the TIRR and LFRR algorithms. Here we can see that the content-based TIRR improves upon the collaborative filtering-based LFRR.}
\label{tbl:collab-f1-score}
\vspace{-4mm}
\end{table}

Table \ref{tbl:collab-f1-score} lists the peak performance metrics for the two algorithms. In addition to the higher F1 score, \textit{TIRR} also has a comparable balance of precision and recall to \textit{LFRR}.

\section{Conclusions}

In this paper, we presented a novel algorithm to interpret user preference history using \textit{only} photos and make predictions about future preferences for reciprocal recommendation. We demonstrated that this can effectively be used as a predictor for the probability of mutual preference between two users, and therefore forms the basis for an effective recommender system. We also demonstrated that our algorithm outperforms state of the art reciprocal recommender systems in offline tests using a large dataset from a dating service with real users.

This research demonstrates the value of including historical preference in reciprocal recommendation. Previous research has demonstrated the value of using RNNs to interpret sequences of preferences in user-item recommendation, but this is the first time it has been used in reciprocal recommendation. The improvement over a similar algorithm that does not use sequences of data shows the value of this approach.

Finally, the model itself represents a significant advance in the field of content-based reciprocal recommendation. The model's success also allows us to draw interesting conclusions about the significance of photos in online dating, given their strong predictive power in this dataset. It also provides interesting insight into the potential power of content-based algorithms in online dating: while in many fields, they are outperformed by collaborative filtering, the algorithm presented in this paper performs better on evaluation metrics than the current state-of-the-art collaborative filtering algorithm.


\bibliographystyle{ACM-Reference-Format}
\bibliography{main}










\end{document}